\documentclass[a4paper,12pt]{article}
\usepackage{fd}
\usepackage[utf8x]{inputenc}

\usepackage{hyperref}    % Hyperlinks in references
\usepackage[all]{hypcap} % Internal hyperlinks to floats.

\usepackage{booktabs}

\usepackage{xcolor} % [dvipsnames] 
\definecolor{halfgray}{gray}{0.55} % chapter numbers will be semi transparent .5 .55 .6 .0
\definecolor{webgreen}{rgb}{0,.5,0}
\definecolor{webbrown}{rgb}{.6,0,0}
\definecolor{Maroon}{cmyk}{0, 0.87, 0.68, 0.32}
\definecolor{RoyalBlue}{cmyk}{1, 0.50, 0, 0}

\usepackage{xspace}
\usepackage{graphicx}
\usepackage{amsmath}
\usepackage{amssymb}
\usepackage{booktabs}
\usepackage{hyperref}
\usepackage{comment}

\def\be{\begin{equation}}
\def\ee{\end{equation}}
\def\beq{\begin{eqnarray}}
\def\eeq{\end{eqnarray}}
\def\bsmumu{\ensuremath{B^0_s \to \mu^{+} \mu^{-}}\xspace}
\def\bsf{\ensuremath{B_s \to f }\xspace}

\def\b0smumu{\ensuremath{B^0_s \to \mu^{+} \mu^{-}}\xspace}

\def\BF{\ensuremath{\mathcal{B}}\xspace}
\def\adeltagammamumu{\ensuremath{\mathcal{A}_{\Delta \Gamma}^{\mu \mu}}\xspace}
\def\adeltagamma{\ensuremath{\mathcal{A}_{\Delta \Gamma}}\xspace}
\def\adeltagammaf{\ensuremath{\mathcal{A}^f_{\Delta \Gamma}}\xspace}
\def\Aa{\ensuremath{\mathcal{A}_{\rm{a}}}\xspace}

\def\Bs{\ensuremath{B_s}\xspace}
\def\Bsz{\ensuremath{B^0_s}\xspace}
\def\Bszbar{\ensuremath{\bar{B_s^0}}\xspace}
\def\Bd{\ensuremath{B_d}\xspace}

\def\effexp{\ensuremath{\varepsilon_{\rm exp}}\xspace}
\usepackage{xcolor} % [dvipsnames] 
\definecolor{webgreen}{rgb}{0,.5,0}
\definecolor{webbrown}{rgb}{.6,0,0}
\definecolor{RoyalBlue}{rgb}{0,0,.5}
\definecolor{halfgray}{gray}{0.55} % chapter numbers will be semi transparent .5 .55 .6 .0
\definecolor{Maroon}{cmyk}{0, 0.87, 0.68, 0.32}

\newcommand{\mc}{\mathcal}

\usepackage{authblk}
\def\mytitle{On the model dependence of measured \Bs-meson branching fractions\xspace}
\hypersetup{%
    colorlinks=true, linktocpage=true, pdfstartpage=1, pdfstartview=FitV,%
    breaklinks=true, pdfpagemode=UseNone, pageanchor=true, pdfpagemode=UseOutlines,%
    plainpages=false, bookmarksnumbered, bookmarksopen=true, bookmarksopenlevel=1,%
    hypertexnames=true, pdfhighlight=/O,%nesting=true,%frenchlinks,%
     urlcolor=webbrown, linkcolor=RoyalBlue, citecolor=webgreen, pagecolor=RoyalBlue,%
    pdftitle={\mytitle},%
    pdfauthor={Francesco Dettori, Diego Guadagnoli},%
    pdfsubject={},%
    pdfkeywords={},%
    pdfcreator={pdfLaTeX},%
}
\title{\mytitle}
\date{}
\author[1]{Francesco Dettori}
\author[2]{Diego Guadagnoli}
\affil[1]{Oliver Lodge Laboratory, University of Liverpool, Liverpool, UK}
\affil[2]{Laboratoire d'Annecy-le-Vieux de Physique Th\'eorique UMR5108\,, Universit\'e de Savoie Mont-Blanc et CNRS, B.P.~110, F-74941, Annecy-le-Vieux Cedex, France}

\begin{document}

\setlength{\parindent}{1.2ex}
\setlength{\parskip}{0.8ex}
\begin{raggedright}
\hfill LAPTH-013/18\\
\end{raggedright}
{\let\newpage\relax\maketitle}

\begin{abstract}

\noindent The measurement of \Bs-meson branching fractions is a fundamental
tool to probe physics beyond the Standard Model. 
Every measurement of untagged time-integrated
\Bs-meson branching fractions is model-dependent due to the time
dependence of the experimental efficiency and the large lifetime
difference between the two \Bs mass eigenstates. In recent measurements, 
this effect is bundled in the systematics.
We reappraise the potential numerical impact of this effect -- we find it to 
be close to 10\% in real-life examples where new physics is a correction to 
dominantly Standard-Model dynamics. 
We therefore suggest that this model dependence be made explicit, 
i.e. that \Bs branching-fraction measurements be presented in a two-dimensional 
plane with the parameter that encodes the model dependence.
We show that ignoring this effect can lead to over-constraining the couplings of 
new-physics models. In particular, we note that the effect also applies when 
setting upper limits on non-observed $\Bs$ decay modes, such as those forbidden 
within the Standard Model.
\end{abstract}

\noindent {\bf Introduction --} The branching fractions of \Bs mesons
belong to the most sensitive probes of physics beyond the Standard Model
(SM) in low-energy, high-intensity experiments.
Their precise measurement is of prime importance to establish
possible new physics or else to constrain models beyond the SM.
However, the comparison between measurements and theory predictions
of \Bs-meson branching fractions presents some subtleties due to the sizeable
lifetime difference $\Delta \Gamma_s$ between the two mass eigenstates of the
$\Bs^0-\bar{\Bs^0}$ system~\cite{DeBruyn:2012wj}.
First of all, in the absence of flavour tagging the measured branching fraction will
be the average of the \Bsz and \Bszbar branching fractions, due to their fast
mixing.
Secondly, since the theoretically calculated branching fraction is usually
defined as the $CP$ average between the flavour eigenstates {\em before any
oscillation}, a $\Delta \Gamma_s$-dependent correction is required for it 
to be compared to the experimental 
values~\cite{DeBruyn:2012wj,DescotesGenon:2011pb}. Both effects are proportional 
to a model- and channel-dependent factor known as 
$\adeltagammaf$ ($f$ denotes the final state). So, in general, the comparison 
between measurements and theoretical predictions involves an assumption about 
this factor.

A third model-dependent bias is introduced by the non-perfect time 
acceptance of real experiments, again because of the sizeable lifetime difference 
$\Delta \Gamma_s$. This effect is discussed in \cite{Knegjens:2014zva}, where it is 
quantified as a 1-3\% correction.\footnote{The effect is also mentioned 
in \cite{DeBruyn:2012wj}~(see sec. V). In the specific context of the \bsmumu
measurement~\cite{Aaij:2012di}, this effect was subsequently developed in Ref.~\cite{Perrin-Terrin:2013cmm} 
and by one of the authors.} In experimental measurements this effect was first 
appreciated in Ref.~\cite{Aaij:2011aj} (see also Ref.~\cite{Aaij:2012di}), 
and in recent results this model-dependent correction is accounted for in the systematic error.

Aim of the present paper is twofold: {\em (i)} we reappraise the relevance of
this effect with respect to existing literature, as we find an O(7\%) correction 
in a realistic example. We accordingly advocate that experiments report explicitly 
the correlation of the result with the value
of the model-dependent parameter ($\adeltagammaf$, or any other parameter
correlated with it),  even when the effect is smaller than the statistical
uncertainty; {\em (ii)} we emphasise that this effect has implications when
setting bounds on new-physics couplings, especially in decay modes where new
physics is not a correction, but the bulk of the dynamics. In such
cases, not properly tracking this effect may even lead to constraints that
qualitatively depart from the dynamics actually at play, as we discuss in a
specific example related to present-day anomalies in flavour data.

We begin by shortly reviewing the basic observation in 
Ref.~\cite{DeBruyn:2012wj}. One starts from the time-dependent untagged decay 
rate for a \Bs into a final state $f$, defined as \cite{Dunietz:2000cr}
\begin{eqnarray}
 \langle \Gamma(B_s (t) \to f) \rangle \equiv \Gamma(B^0_s (t) \to f)+
\Gamma(\bar B^0_s (t) \to f) = R_H^f e^{-\Gamma_H t} +R_L^f e^{-\Gamma_L t} =
\nonumber \\
  = (R_H^f + R_L^f ) e^{-\Gamma_s t } \left[ \cosh \left( \frac{y_s t
}{\tau_{B_s}} \right) + \adeltagammaf \sinh \left(\frac{y_s t }{\tau_{B_s}}
\right)\right]~,
  \label{eq:timedep}
\end{eqnarray}
where, in standard notation \cite{PDG2016}, $\Gamma_s = 1/\tau_{B_s}$
is the average between the widths, $\Gamma_H$ and $\Gamma_L$, of the two mass
eigenstates in the \Bs system. 
The parameter $y_s =\frac{\Gamma_L - \Gamma_H}{2\Gamma_s} = \frac{\Delta \Gamma_s }{2 \Gamma_s}$
quantifies the generic size of effects due to the \Bs-system width difference, $y_s = 0.061(4)$ \cite{Amhis:2016xyh}. 
Finally $\adeltagammaf = \frac{R^{f}_H - R^f_L}{R^{f}_H + R^f_L}$
depends on the final state and is related to the underlying
dynamics, hence being model-dependent. The time-integrated branching ratio is then
obtained by integrating eq. (\ref{eq:timedep}):
\begin{equation}
\label{eq:brexp}
 \BF_{\rm ave} (\bsf) = \frac{1}{2} \int_0^\infty \langle \Gamma (B_s(t) \to f) \rangle d t 
= ( R^{f}_H + R^f_L ) \frac{\tau_{\Bs}}{2} \left[\frac{1+ \adeltagammaf \, y_s}{1-y_s^2} \right]~.
\end{equation}
As noted in Ref.~\cite{DeBruyn:2012wj}, this is different from the theoretical
branching fraction, which is usually calculated as $CP$-averaged at time zero:
\begin{equation}
\label{eq:brth}
\BF_{\rm th}(B_s \to f) \equiv \frac{\tau_{B_s}}{2} \langle \Gamma(B_s (t) \to f) \rangle|_{t=0}~,
\end{equation}
so that even with a perfect experiment, a model-dependent correction is needed
to compare with the time-integrated branching fraction, $\BF_{\rm ave}$: 
\begin{equation}
\label{eq:br_th_exp}
 \BF_{\rm th} (\bsf) = \left(\frac{1 - y^2_s}{1 + \adeltagammaf \, y_s }\right) \BF_{\rm ave} (\bsf)~. 
\end{equation}

\noindent {\bf Time-dependent efficiencies --} However, experiments are not
perfect. In particular, the integral of the rate over the meson proper time is
sampled according to a time-dependent efficiency. Hence, the experimentally
measured branching fraction is actually
\begin{equation}
\label{eq:brreal}
 \BF_{\rm exp} (\bsf) = \frac{N_{\rm{obs}}}{N \effexp} =  
\frac{1}{2 \effexp}\int_0^\infty  \varepsilon(t) \langle\Gamma (B_s(t) \to
f) \rangle d t
\end{equation}
where $\varepsilon(t)$ is the time-dependent efficiency of the apparatus,
\effexp is the time-averaged efficiency with which the observed
yield, $N_{\rm{obs}}$, is corrected, and $N$ is the total number of mesons produced
to which the experiment normalises. 

Unless $\varepsilon(t)$ is perfectly constant, the apparatus efficiency
introduces an extra dependence on $\adeltagammaf$, and the latter makes the
measurement of eq.~\eqref{eq:brreal} model dependent. This dependence cannot be
factorised and accounted for as in eq.~\eqref{eq:br_th_exp} as it rests on the
explicit functional form of the efficiency. Intuitively, the rates of the two
physical eigenstates will not be sampled uniformly, and this will distort the
more the physical decay distribution, the more the two lifetimes differ.
As a consequence, the measured admixture is not as given by the r.h.s. of eq.
(\ref{eq:brexp}), and the dependence on $\adeltagammaf$ in the relation between
the calculated and the measured branching fraction is not as simple as given in
eq. (\ref{eq:br_th_exp}).

This bias could be simply corrected for if $\adeltagammaf$ could be univocally fixed
for each given decay channel $f$. However \adeltagammaf depends on the
short-distance structure of the decay, 
hence it is in general different in models of new physics with respect to the SM. 
For example, within the SM for the $\Bs \to \mu^+ \mu^-$ decay one has $\adeltagammamumu = +1$, 
i.e. that the decay occurs mostly through the heavier $\Bs$ eigenstate ($R_L = 0$) \cite{DeBruyn:2012wk}.
This assumes negligible $CP$ violation in mixing and in the
interference between decays with and without mixing --
an assumption that turns out to be robust.
However, the  \bsmumu decay could receive contributions beyond the SM 
from semileptonic scalar and pseudoscalar couplings,  
whose current bounds do not actually exclude any $\adeltagammamumu$
value in the whole range $[-1, +1]$~\cite{DeBruyn:2012wk,Buras:2013uqa}. 

One clear way to expose the measurements' dependence on the value of
\adeltagammaf, and the ensuing model dependence would be to present measurements
as a function of the assumed value for \adeltagammaf. Of course, such practice
is not always necessary. Notably, if the mixture of the heavy and light
eigenstates is known for a given final state, the effect can be properly
accounted for in the experimental efficiency. For example, $\adeltagammaf = 0$
for flavour-specific decays. Furthermore, this effect is diluted or absent in
decay rates where the SM contribution is precisely known and dominant. This
effect can instead be prominent in rare decays, whose branching fractions can
receive large contributions from new physics. We now illustrate such effect with
a concrete example (see also \cite{Knegjens:2014zva}).

While the functional form of the time-dependent efficiency can be non-trivial,
to estimate the size of the bias one may assume a simple step function $\varepsilon(t) = \theta(t - t_0)$, 
i.e. $\varepsilon = 0$ for $t<t_0$ and $\varepsilon=1$ elsewhere. With this function one gets 
\begin{eqnarray}
&&\frac{1}{2}\int_{0}^\infty  \varepsilon(t) \langle\Gamma (B_s(t) \to f) \rangle d t = \nonumber \\
&& ( R^f_H + R^f_L ) \frac{\tau_{\Bs}}{2} \frac{e^{- \Gamma_s t_0}}{1-y_s^2} 
\left[ \cosh\left(  \Gamma_s y_s \, t_0 \right) (1 + \adeltagammaf \, y_s) + 
\sinh\left(  \Gamma_s y_s \, t_0 \right) (y_s + \adeltagammaf)\right]~,~~~~
\end{eqnarray}
which clearly reduces to eq.~(\ref{eq:brexp}) for $t_0 = 0$. One can accordingly
define the bias $\delta$ with respect to the branching ratio obtained with
constant efficiency as the function
\begin{equation}
\nonumber
\delta(\adeltagammaf, y_s, \effexp) \equiv \frac{\BF_{\rm exp}(\Bs \to f)}{\BF_{\rm ave}(\Bs \to f)}
= \frac{e^{- \Gamma_s t_0}}{\effexp} 
 \left( \cosh\left( \Gamma_s y_s \, t_0  \right) + 
\sinh\left( \Gamma_s y_s \, t_0\right) 
\frac{y_s + \adeltagammaf}{1+ \adeltagammaf \, y_s} \right)~,
\end{equation}
where the efficiency correction appears explicitly as in eq.~\eqref{eq:brreal}. 
This efficiency is estimated by making a definite assumption about \adeltagammaf, namely as
\begin{equation}
\effexp(\mathcal{A}_{\rm{a}}) =  \frac{\int_{0}^\infty  \varepsilon(t) \langle\Gamma_{\rm{a}} (B_s(t) \to f \rangle d t }
{\int_{0}^\infty  \langle\Gamma_{\rm{a}} (B_s(t) \to f \rangle d t }
\end{equation}
where $\Gamma_{\rm{a}}$ is the time-dependent width under the assumption $\adeltagammaf = \Aa$. 
Here we posit that the experimenter can estimate $\varepsilon(t)$ with good accuracy 
from auxiliary measurements, typically from control channels, or else from Monte Carlo simulations.
\begin{figure}[!t]
\centering \includegraphics[width = 0.6\textwidth]{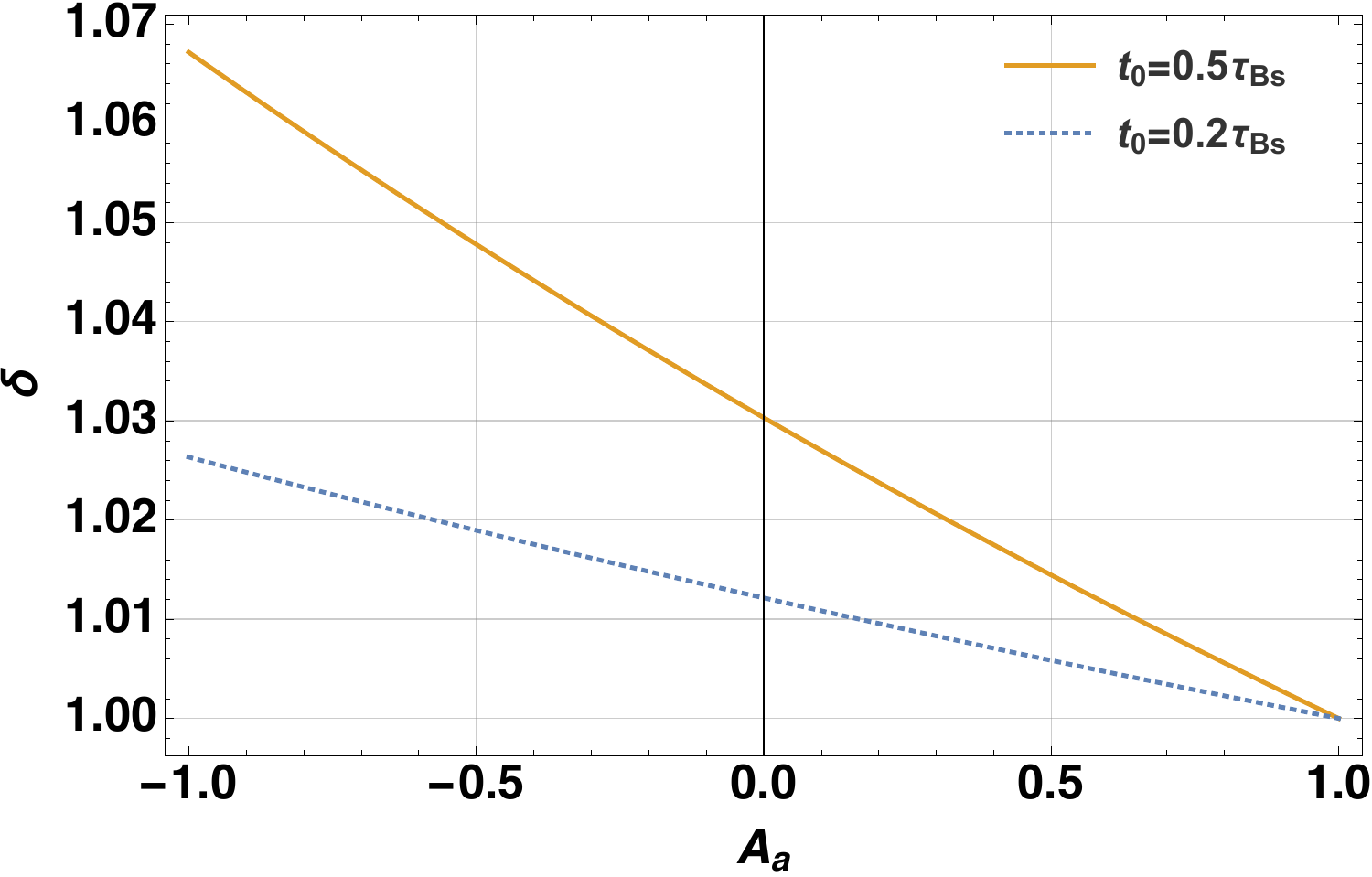}
 \caption{The bias $\delta$ as a function of the assumed value for
$\adeltagammaf$, \Aa, for a decay with $\adeltagammaf=1$. The efficiency
function is modelled as a step function $\theta(t - t_0)$, with two realistic
$t_0$ values.}
\label{fig:delta}
\end{figure}
The bias will be therefore a function of \Aa: 
\begin{equation}
 \delta(\adeltagammaf, y_s, \Aa) = 
 \frac{ \cosh\left( \Gamma_s y_s \, t_0  \right) + 
\sinh\left( \Gamma_s y_s \, t_0\right) 
\frac{y_s + \adeltagammaf}{1+ \adeltagammaf \, y_s} }{ 
\cosh\left( \Gamma_s y_s \, t_0  \right) + 
\sinh\left( \Gamma_s y_s \, t_0\right) 
\frac{y_s + \Aa}{1+ \Aa \, y_s}}
\end{equation}
which is by construction equal to 1 when the assumed value \Aa for $\adeltagammaf$ 
coincides with the physical one. 
Hence in practice \effexp has to be calculated for each value of \Aa, so that for
the same experimental event yield the branching fraction can be properly
estimated for an assumed model.
We illustrate the numerical impact of the bias $\delta$ in Fig.~\ref{fig:delta}. 
Here $\delta$ is shown as a function of $\Aa$, under the hypothesis that the physical $\adeltagammaf=1$, 
and for two realistic values of $t_0$. 
In this example the bias amounts to overestimating the measured branching fraction 
with respect to the real one: as soon as the assumed value of $\adeltagammaf$, \Aa, 
departs from the physical value, the bias $\delta$ is larger than 1. 
This is as expected. 
In fact, with the considered efficiency function,
estimating $\varepsilon_{\rm exp}$ with $\Aa < +1$ means that one
is undersampling the heavy eigenstate, the only one actually
contributing if the physical $\adeltagammaf = +1$.
As a consequence, $\varepsilon_{\rm exp}$ in eq. (\ref{eq:brreal}) 
is smaller than the correct value that one would obtain for the physical $\adeltagammaf = +1$.
As the figure shows, for values as low as $t_0 = 0.5 \tau_{\Bs}$ the bias can be as large as $\sim 7\%$. 

Conversely, if one assumes that the inefficiency is for high proper-time values,
 \mbox{$\varepsilon(t) = \theta(t_0-t)$}, then the bias will be in the opposite direction. 
In general, in real experiments one can expect inefficiencies both at low and at high proper-time values, 
so that the convolution with the expected time distribution will be performed by means
of Monte Carlo simulations. 

\begin{figure}[!t]
\centering 
 \includegraphics[width = 0.6\textwidth]{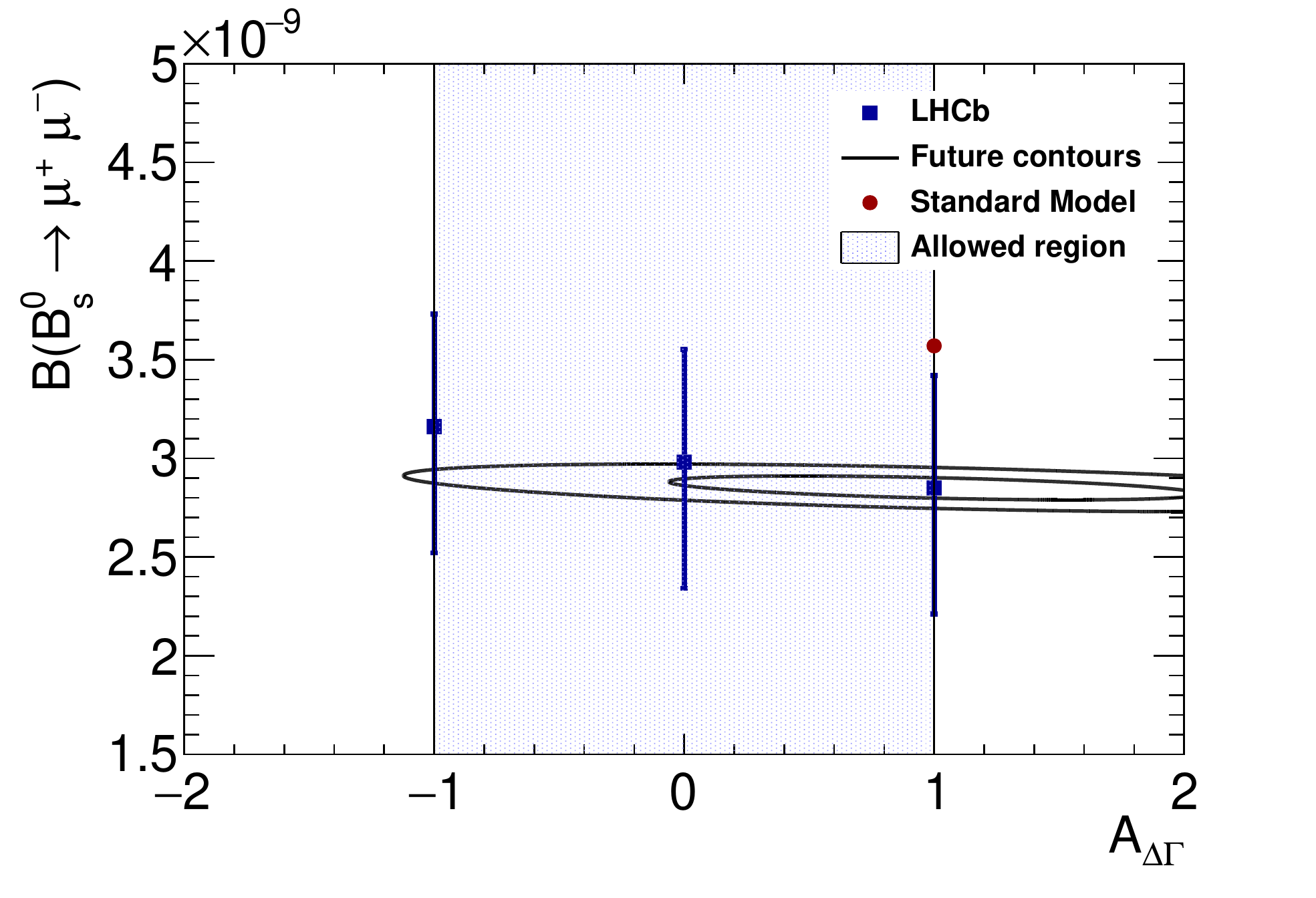}
\caption{\small 
LHCb measurement of the \bsmumu branching fraction vs. $\adeltagammamumu$ (blue
squares)~\cite{Aaij:2017vad}. The respective SM predictions are also reported
(red circle). Black ellipses show 1- and 2-$\sigma$ contours of a possible
future measurement of the two observables simultaneously (see text).
}\label{fig:adeltagamma}
\end{figure}

\noindent {\bf Current status --} In the majority of recent \Bs
branching fraction measurements, the effect of the possible model dependence
generated by a time-dependent efficiency has been treated as a systematic uncertainty, 
e.g. see Refs.~\cite{Aaij:2015esa,Aaltonen:2011sy,Aaij:2015uoa,Aaij:2017zpx}. 
On the other hand, only in very
few examples is the effect treated as full-fledged dependence -- which is what we advocate. 
An example of such treatment is the
latest LHCb measurement of $\BF(\bsmumu)$~\cite{Aaij:2017vad}, where
the branching fraction is quoted for the SM assumption ($\adeltagammaf = 1$),
and corrections for $\adeltagammaf = \{0, -1\}$ are reported. 
The size of the variation is respectively +4.6\% ($\adeltagammaf = 0$) and
+10.9\% ($\adeltagammaf = -1$). 
This is displayed in Fig.~\ref{fig:adeltagamma} where the three values are shown
in the two-dimensional plane of branching fraction and \adeltagammaf, together with the
SM prediction~\cite{Beneke:2017vpq}. 
We also note that Ref.~\cite{Aaij:2017vad} reports a measurement of the \bsmumu effective
lifetime ($\tau_{\mu\mu}$)~\cite{Fleischer:2010ib,Fleischer:2011cw,DeBruyn:2012wk}, which 
is in turn directly sensitive to \adeltagammamumu itself. Therefore the two 
observables could already be represented in a two-dimensional plane, although 
the current $\tau_{\mu\mu}$ measurement would translate into 
$\adeltagammamumu = 8 \pm 11$, whose central value lies in the non-physical 
region but with large uncertainty. An illustrative example of such a correlated 
measurement is again in Fig.~\ref{fig:adeltagamma}. In particular, the lines 
labelled ``future contours'' represent 1- and 2-$\sigma$ contours assuming the 
current central value of the branching fraction with $\adeltagammamumu =1$, and 
a tenfold smaller uncertainties with respect to the LHCb 
measurement~\cite{Aaij:2017vad}.

\begin{figure}[t]
\centering \includegraphics[width = 0.6\textwidth]{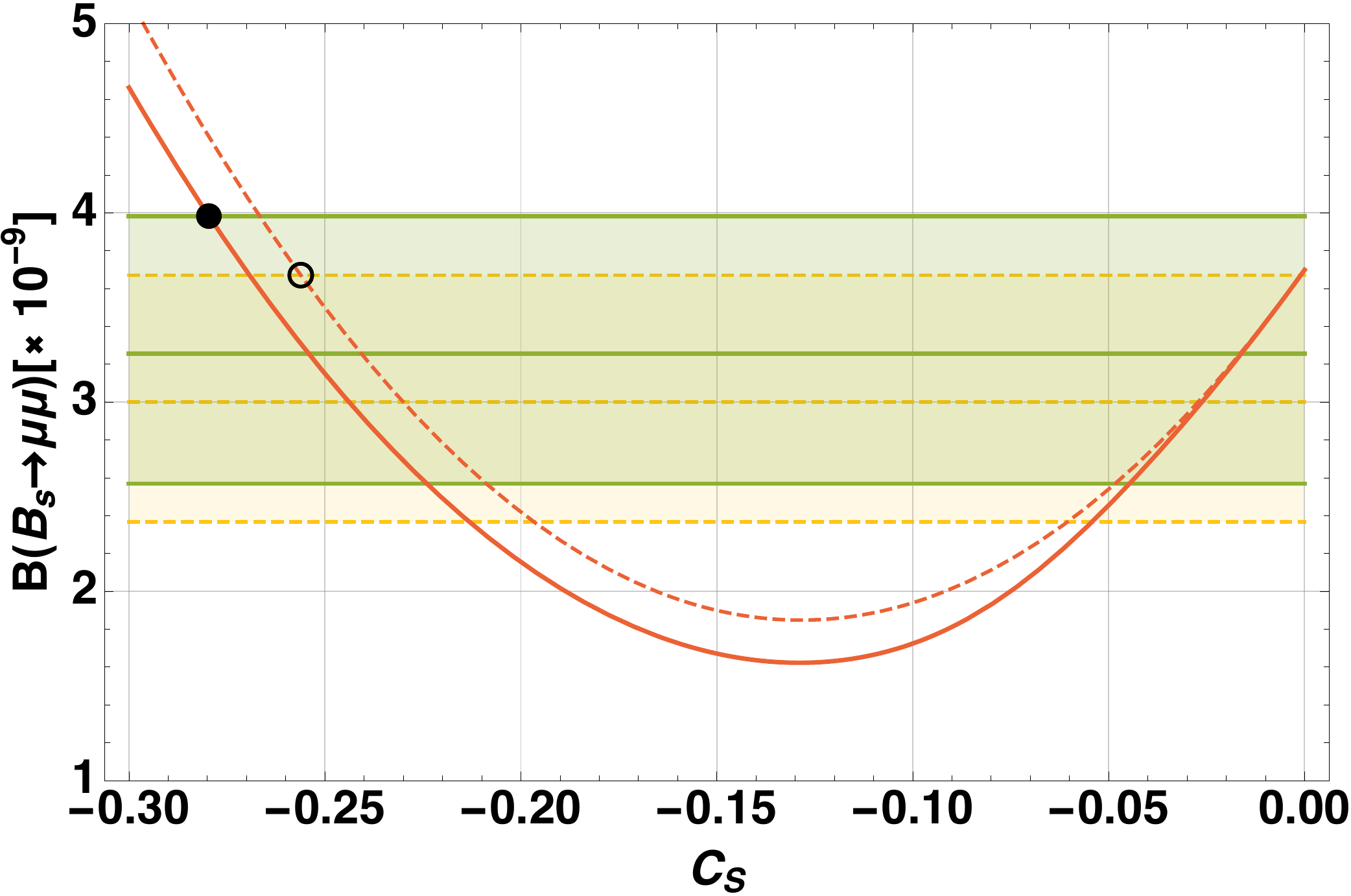}
 \caption{Red lines: theory predictions as a function of a scalar Wilson-coefficient 
shift $C_S = -C_P$, for $\adeltagammamumu = +1$ (dashed) 
and respectively $\adeltagamma(C_S)$ (solid). 
Horizontal bands: experimental ranges for $\adeltagammamumu = +1$ (yellow dashed), 
and respectively $\adeltagammamumu (\bar C_S)$, where $\bar C_S$ corresponds to the filled dot in the figure. 
See text for more details.}
\label{fig:BRvsCS}
\end{figure}

\noindent {\bf Biases on the Wilson coefficients --}
Neglecting the discussed variation can lead to an over-constraining of the theory
parameter space, notably in models with sizeable scalar or pseudo-scalar
contributions (with arbitrary phases), as illustrated by the following example.
Let us consider a shift to the Wilson coefficients $C_{S,P}$ of the operators
\be
\label{eq:OS_OP}
\mc O_S = \frac{e^2}{16 \pi^2} (\bar s P_R b) (\bar \ell \ell)~,~~~~~~~~ \mc O_P = \frac{e^2}{16 \pi^2} (\bar s P_R b) (\bar \ell \gamma_5 \ell)~,
\ee
that can give sizeable contributions to the \bsmumu rate. 
Let us assume they fulfil the constraint $C_S = - C_P$, as generally expected
for new physics above the electroweak symmetry-breaking scale
\cite{Alonso:2014csa}. The $\mc B(\bsmumu)$ prediction as a function of
$C_S$, and corrected by the factor $(1 + \adeltagammaf \,
y_s)/(1 - y^2_s)$ (see eq. (\ref{eq:br_th_exp})), is displayed in Fig.
\ref{fig:BRvsCS} for two choices of $\adeltagammamumu$. The first choice is
$\adeltagammamumu = +1$, shown as a red dashed curve. The latest LHCb
measurement corresponding to this value of $\adeltagammamumu$ is shown as a
yellow dashed horizontal band. The upper line of this band and the red dashed
curve intersect at $C_S \simeq - 0.25$ which may be taken as a $1\sigma$ bound
on $C_S$. However, $\adeltagammamumu = \adeltagammamumu(C_S)$
\cite{DeBruyn:2012wk}: the theory prediction corrected for this dependence,
again through the $(1 + \adeltagammaf(C_S) \, y_s)/(1 - y^2_s)$ factor, is
displayed as a solid red curve. Concurrently, also the experimental measurement
is a function of $\adeltagammamumu$ as we have discussed. In the figure we show
as a solid green band the measurement for $\adeltagammamumu = -0.56$, which
corresponds to $C_S \simeq -0.28$, the value at which the theory prediction and
the experimental central value +1$\sigma$ intersect. It is this $C_S$ value that
should be taken as the correct $1\sigma$ bound on $C_S$. We see that the
difference between the two bounds, obtained respectively for $\adeltagammamumu =
+1$ and the correct $\adeltagammamumu$, is of O(10\%).

Of course, the size of the effect just described will depend on the
relative importance of scalar operators in the process being constrained. 
While intuitively the size $\lesssim$ O($10\%$) of the experimental bias 
-- concretely, the variation of the branching-ratio measurement with 
$\adeltagammaf$ -- is expected to provide an upper bound on the size of the 
corresponding bias on Wilson coefficients, we would like to put forward an 
example where the latter bias turns out to be larger. This example is relevant 
in view of the existing discrepancies in flavour physics, and underlines the 
necessity of precisely tracking the theory that is being constrained (hence 
assumed), as soon as the measured $\adeltagammaf$ in a given decay mode $B_s \to 
f$ should differ from the assumed one. This in turn highlights the importance of 
effective-lifetime measurements, pointed out in 
\cite{Fleischer:2010ib,Fleischer:2011cw,DeBruyn:2012wk}, that are a probe of 
$\adeltagammaf$. Let us consider the effective-theory description emerging from 
present-day discrepancies in $b \to s \mu \mu$ data, in particular by the lepton 
universality violation (LUV) tests $R_K$ and $R_{K^*}$ measurements 
\cite{Aaij:2014ora,Aaij:2017vbb}. 
Among the preferred explanations in terms of shifts to the Wilson coefficients 
of the $b \to s$
effective Hamiltonian, an important one is the scenario with opposite 
contributions to the
operators $\mathcal O_9 \propto (\bar s \gamma^\alpha_L b) \, (\bar \mu
\gamma_{\alpha} \mu)$ and $\mathcal O_{10} \propto (\bar s \gamma^\alpha_L b) \,
(\bar \mu \gamma_{\alpha} \gamma^5 \mu)$. 
In particular a shift $\delta C_9^\mu = - 
\delta C_{10}^\mu \simeq - 13\% |C_{10}^{\rm SM}| \approx -0.5$ to the 
$C^{\mu}_{9(10), \rm SM}$ Wilson coefficients is preferred~\cite{Hiller:2014yaa,Ghosh:2014awa}. 
The structure resulting from such shifts, 
$(\bar s \gamma^\alpha_L b) \, (\bar \mu \gamma_{\alpha L} \mu)$, has a 
$(V-A) \times (V-A)$ form and as such is very suggestive from the point 
of view of the ultraviolet dynamics, e.g. it can be straightforwardly rewritten 
in terms of $SU(2)_L$-invariant fields 
\cite{Alonso:2014csa,Bhattacharya:2014wla}. Since the effective scale of such 
structure lies typically above the electroweak scale, the fermion fields
involved will in general not be aligned with the mass basis. Hence, below the electroweak symmetry-breaking scale, such 
structure, introduced to account for LUV, will also generate lepton flavour 
violating dynamics, whose size is related to the {\em measured} amount of LUV~\cite{Glashow:2014iga}. From this argument, 
the analogous $(V-A)\times (V-A)$ operator $(\bar s \gamma^\alpha_L b) \, (\bar 
\ell \gamma_{\alpha L} \ell^\prime)$ would contribute to processes such as $B_s 
\to \ell^- \ell^{\prime +}$, if a similar structure with the appropriate flavour 
indices is also favoured to explain LUV.
Such argument does not forbid contributions from scalar operators of 
comparable size. Actually, constraints on scalar contributions (for recent 
analyses see \cite{Altmannshofer:2017wqy,Fleischer:2017ltw}) are substantially 
weakened to the extent that a shift to $C_{10}$ is at play, as we discuss 
next.\footnote{
Sensitivity of rare decays to scalar operators is warranted by the fact
that the fermion mass necessary to perform the chiral flip may actually
be a large mass, at variance with the SM case. Sizeable scalar
contributions are accordingly ubiquitous as soon as the bosonic sector
is enlarged with respect to the sheer SM content.}
In any of the $B_s \to \ell^- \ell^{\prime +}$ decays, contributions 
from the Wilson coefficients of the operators
\beq
\label{eq:O_LFV}
\begin{tabular}{ll}
$\mc O_9^{\ell \ell'} \equiv \frac{e^2}{16 \pi^2}(\bar s \gamma^\alpha_L b) \, (\bar \ell \gamma_{\alpha} \ell^\prime)~,$ & 
$\mc O_{10}^{\ell \ell'} \equiv \frac{e^2}{16 \pi^2}(\bar s \gamma^\alpha_L b) \, (\bar \ell \gamma_{\alpha} \gamma_5 \ell^\prime)$~,\\
[0.2cm]
$\mc O_S^{\ell \ell'} \equiv m_b \frac{e^2}{16 \pi^2}(\bar s P_R b) \, (\bar \ell \ell^\prime)~,$ &
$\mc O_P^{\ell \ell'} \equiv m_b \frac{e^2}{16 \pi^2} (\bar s P_R b) \, (\bar \ell \gamma_5 \ell^\prime)~,$\\
\end{tabular}
\eeq
are of the form (see e.g. \cite{Guadagnoli:2016erb})
\be
\mc B(B_s \to \ell^+_1 \ell^-_2) \propto 
(1 - \hat{m}^{2} )|F_{P} + \hat{M} C_{10}|^{2} +
(1 - \hat{M}^{2} )|F_{S} - \hat{m} C_{9}|^{2}~,
\ee
where $\hat{m} \equiv \hat m_{\ell_2} - \hat m_{\ell_1}$, $\hat{M} \equiv \hat 
m_{\ell_1} + \hat m_{\ell_2}$, with hats denoting that the given mass is 
normalized by $M_{B_s}$, and where $F_{S, P} \approx M_{B_{s}} C_{S, P}$.
A sizeable departure in $\adeltagammaf$ from unity would signal accordingly 
sizeable contributions from $C_{S,P}$. In particular, $C_P$ could partly cancel 
(depending on its phase, which is unconstrained) the contribution from $C_{10}$ 
so that the measured signal would actually be due to $C_S$ dominantly, and this 
is the Wilson coefficient that the measurement would constrain in reality.
In these circumstances, if one insisted with the assumption
$\adeltagammaf = +1$, one would, instead, interpret the branching-ratio
measurement as a constraint to $C_{10}$, under the hypothesis that scalar
contributions are negligible. So, the combination of Wilson coefficients that
is actually constrained by a $B_s \to f$ decay measurement needs be carefully
tracked as soon as $\adeltagammaf$ is measured and departs from
unity.\footnote{We emphasise that our argument holds for LU and lepton-flavour
conserving decays alike.}

In short, it will be important to present future experimental 
measurements in a two-dimensional plane of the branching fraction and either 
\adeltagammaf or another observables correlated with it, such as the effective
lifetime. A quite useful example is Ref. \cite{Aaij:2017cza}, where the limit is
quoted for $\adeltagammaf = \{-1, 1\}$, thus allowing a handy extrapolation
to any scenario with shifts to the operators in the second line of eq.
(\ref{eq:O_LFV}). 

\noindent {\bf Other considerations --} It is clear that if time
information is available and the statistics are sufficient to perform a
time-dependent analysis, the effect described in this paper is no longer present
as the time-dependent efficiency can be convoluted with the correct time
distribution. Secondly, this effect is even more relevant when combining
different experimental measurements, as different apparatuses can have a
different time-dependent efficiency and thus a different dependence on
\adeltagammaf. In third place, since this effect depends experimentally on the
apparatus efficiency {\em and not on the yield}, it is also present when setting
limits on branching fractions; for example, it does apply to limits on channels
forbidden in the SM and, as we argued, it may be a large effect there.

Finally, we note that this effect was presented here for the case of \Bs mesons but in fact it is more general. 
The measurement of a branching fraction of a meson that oscillates is model dependent if 
\begin{enumerate}
 \item the experiment is realistic, i.e. $\varepsilon(t)$ is not constant over the whole proper-time range;
 \item the final state $f$ is available to both mass eigenstates;
 \item the difference in lifetime between the mass eigenstates is not negligible with respect to the meson average lifetime.
\end{enumerate}
In practice the last condition is realized only for \Bs mesons so far.
In fact, while for \Bs mesons $\Delta \Gamma_s$ is sizeable compared to $\Gamma_s$, 
this is not true for \Bd or $D^0$ mesons. 
In the other relevant case of $K^0$ mesons, the difference in
lifetimes between $K_S$ and $K_L$ is so large that branching fractions are
directly reported for the two mass eigenstates rather than for the flavour
ones. If one had to report branching fractions for the $K^0$ and $\bar K^0$ the effect
here described would be maximal.

\noindent {\bf Summary --} Every measurement of a \Bs untagged time-integrated
branching fraction is model dependent due to the time dependence of the
experimental efficiency \cite{Knegjens:2014zva,DeBruyn:2012wj}. We show with two
real-life examples that this dependence can be as large as O(10\%), and argue
that it needs be properly tracked. We accordingly suggest that \Bs
branching-fraction measurements be presented in a two-dimensional plane with the
parameter \adeltagammaf or another observable correlated with it, even in the
case the latter would not be yet measurable. We also argue that theoretical
predictions within a given model should be compared with the measured value of
the branching fraction corresponding to the \adeltagammaf value calculated
assuming the same model. These practices should also be carried out for upper
limits on the branching fraction of non-observed channels, notably those
forbidden in the SM, where new physics is dominant, rather than just a
correction. Ignoring this effect may lead to over-constraining new-physics
couplings, or even to constraints that qualitatively depart from the dynamics
actually at play.

\section*{Acknowledgements}

The authors are indebted to Tim Gershon and Patrick Koppenburg for
crucial remarks on the first preprint version of the manuscript. We also 
acknowledge useful comments from Peter Stangl. FD would like 
to thank Marc-Olivier Bettler, Francesca Dordei, Niels Tuning and Tara Shears for
discussion and comments on the manuscript. The work of DG is partially supported
by the CNRS grant PICS07229.
FD acknowledges support from the Science and Technology Facilities Council, UK. 

\bibliographystyle{fd}

\bibliography{main}
\label{biblio}

\end{document}